\documentclass[prd,a4paper,nofootinbib]{revtex4}
\usepackage{mathrsfs}
\usepackage{graphicx}
\usepackage{dcolumn}
\usepackage{bm}
\usepackage{epsfig}
\usepackage{slashed}
\newcommand{\be}{\begin{equation}}
\newcommand{\ee}{\end{equation}}
\newcommand{\bea}{\begin{eqnarray}}
\newcommand{\eea}{\end{eqnarray}}
\newcommand{\bean}{\begin{eqnarray*}}
\newcommand{\eean}{\end{eqnarray*}}

\newcommand{\gapproxeq}{\lower
.7ex\hbox{$\;\stackrel{\textstyle >}{\sim}\;$}}
\newcommand{\lapproxeq}{\lower
.7ex\hbox{$\;\stackrel{\textstyle <}{\sim}\;$}}

\begin{document}

\bibliographystyle{unsrt}

\title{ Lineshape of $e^+ e^-\to D^* \bar D+c.c.$ and electromagnetic form factor of $D^*\to D$ transition in the time-like region}

\author{Yuan-Jiang Zhang$^{1}$\footnote{E-mail: yjzhang@ihep.ac.cn},
and Qiang Zhao$^{1,2}$\footnote{E-mail: zhaoq@ihep.ac.cn}}

\affiliation{1) Institute of High Energy Physics, Chinese Academy of
Sciences, Beijing 100049, P.R. China}

\affiliation{2) Theoretical Physics Center for Science Facilities,
CAS, Beijing 100049, P.R. China}

\date{\today}

\begin{abstract}

In this work, we apply the vector meson dominance (VMD) model to
extract the electromagnetic time-like form factor of the $D^*\to D$
transition combining the recent Belle data for $e^+ e^-\to D^{*+}
D^- + c.c.$ and data for $D^*\to D\gamma$. Two solutions are
obtained in the interpretation of the cross section lineshape: i)
With a relatively large coupling for $\psi D^*\bar{D}$ determined by
experiment, destructive interferences among those charmonium
components are required to bring down the overall cross sections,
and then account for the cross section lineshape. ii) With a
relatively small value for the $\psi D^*\bar{D}$ coupling based on
heavy quark theory, an apparent cross section deficit near threshold
is observed, and contributions from other mechanisms are needed. It
might imply the presence of an additional resonance $X(3900)$.
Meanwhile, we also point out that an enhancement like that could be
produced by the $D_s^*\bar{D_s}+c.c.$ open channel effects.

\end{abstract}

\maketitle





\vspace{1cm}

\section{Introduction}

The $D\bar{D^*}+c.c.$ productions in $e^+e^-$ annihilation give
access to the study of the time-like electromagnetic (EM) form
factor of $D^* \to D\gamma^*$ transition in the charmonium mass
region. Their cross sections were measured recently by
Belle~\cite{Abe:2006fj} and BABAR~\cite{:2009xs}, and clear
resonance structures were observed above the $D\bar{D^*}$ or
$D^*\bar{D}$ threshold. In the real photon limit, the coupling form
factor can be measured via $D^*\to D\gamma$, which turns out to be
an important decay mode for both the charged and neutral $D^*$
mesons~\cite{pdg2008}. In particular, it shows that the partial
decay coupling for $D^{*0}\to D^0\gamma$ could be much larger than
that for $D^{*\pm}\to D^\pm\gamma$. This feature initiated a lot of
efforts on understanding the $D^*\to D$ transition form factor.

Our motivation in this work is to study the $D^*\to D\gamma^*$ form
factor in the time-like region with the help of the recent
experimental data~\cite{Abe:2006fj,:2009xs}. We shall take into
account the resonance contributions to the form factor by employing
the extended vector meson dominance (VMD)
model~\cite{Bauer:1977iq,Bauer:1975bw}. To connect the real photon
limit to the energy region above the $D^*\bar{D}$ threshold, we also
include the light vector meson contributions. In
Ref.~\cite{Aliev:1994qf}, a VMD model was adopted for studying the
$D^*\to D\gamma^*$ form factor. However, due to lack of experimental
information at that time, the authors assumed that the widths for
all vector mesons apart from the $\psi(4040)$ (and beyond) are zero.
This should be a too-rough approximation. As studied recently in
Ref.~\cite{Zhang:2009gy}, the width effects were found essentially
important for understanding the cross section lineshape of $e^+
e^-\to D\bar{D}$.

Another useful and correlated channel is $D^*\to D e^+ e^-$, which
probes the time-like form factor in small momentum squared region.
However, due to the significant suppression of the EM vertex,
branching ratio of this channel is expected to be very small and
hard to measure. This branching ratio can be calculated in our model
and serves as a prediction from theory.

There is a great advantage for extracting the $D^*\to D\gamma^*$
form factor in $e^+ e^- \to D^*\bar{D} +c.c.$ Namely, there is only
one Lorentz structure for the $VVP$ coupling, where $V$ and $P$
stand for vector and pseudoscalar meson fields, respectively.
Therefore, all information about the transition mechanisms would be
contained in a single coupling form factor, which is a complex
function of the photon's four-vector momentum squared. Our
calculations will be compared with the Belle data for $e^+e^-\to D^+
D^{*-}+c.c.$~\cite{Abe:2006fj}.

As follows, we first present the details of the VMD model in
Sec.~\ref{sec-ii}. The numerical results will be given in
Sec.~\ref{sec-iii}. Section~\ref{sec-iv} is devoted to a summary and
discussion.

\section{The model}\label{sec-ii}

The typical effective Lagrangian for the $ \gamma^* D^* \bar D $ and
$\gamma^* \bar{D^*} D$ coupling can be written as:
\be
\mathcal L = -i e g_{\gamma^* D^* \bar D}
\varepsilon_{\alpha\beta\mu\nu}
\partial^\alpha A^\beta
\partial^\mu {D^*}^\nu \bar D + h.c.,\label{lag-1}
\ee
where $A^\beta$ is the vector meson and electromagnetic field,
$\varepsilon_{\alpha\beta\mu\nu}$ is the antisymmetric tensor. With
Eq.~(\ref{lag-1}), the matrix element of $e^+e^- \to D^* \bar D$ in
the one-photon approximation can be written as:
\bea
T = e^2 \bar {v}(k_2) \gamma_\alpha u(k_1) \frac{1}{s} g_{\gamma^*
D^* \bar D}(s)\varepsilon_{\alpha\beta\mu\nu} p_{\bar D}^\beta
p_{D^*}^\mu\epsilon^\nu,\label{element-1}
\eea
where $u(k_1)$ and $v(k_2)$ are the Dirac spinors of the electron
and positron, respectively; $\epsilon^\nu$ represents the
$D^*$-meson polarization vector, and $g_{\gamma^* D^* \bar D}(s)$ is
the effective coupling form factor for the $D^* \to D$ transition.
Note that the electron charge $e$ has been isolated out in this
definition. In the above equation, $s=(k_1 + k_2)^2$ is the overall
center mass energy, while $p_{\bar D}$ and $p_{D^*}$ are the
four-vector momenta of the final state $\bar D$ and $D^*$ meson.

\begin{figure}[ht]
\scalebox{0.5}{\includegraphics{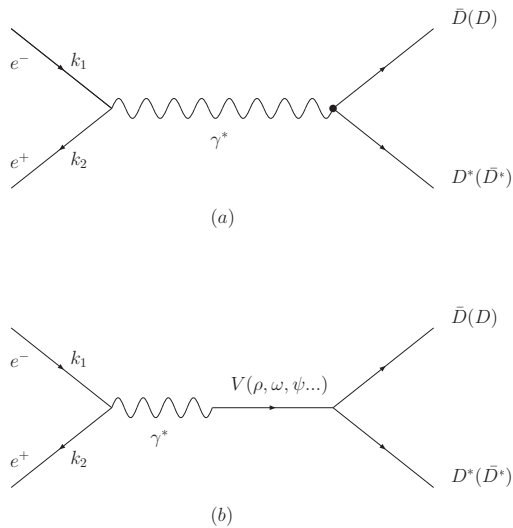}}
\caption{Schematic diagrams based on the VMD model for $e^+e^-\to
D^* \bar D +c.c.$ Diagram (a) is for the single photon approximation
with an effective coupling $g_{\gamma^*D^*\bar{D}}(s)$, while (b)
represents that the electromagnetic field is decomposed into the sum
of vector meson fields.} \label{fig-1}
\end{figure}

As shown in Fig.~\ref{fig-1}, with the VMD
model~\cite{Bauer:1975bw,Bauer:1977iq} we can decompose the
electromagnetic current into a sum of all vector meson fields
including both isospin-0 and isospin-1 components. The $V\gamma^*$
effective coupling can be written as: \be \mathcal L _{V\gamma} =
\sum_V \frac{e M_V^2}{f_V} V_\mu A^\mu, \label{lag-2} \ee where
$V^\mu(=\rho,\omega,\phi, J/\psi ...)$ is the vector meson field,
and $e M_V^2/f_V$ is the photon-vector-meson coupling constant.
Setting $m_e \simeq 0$, $e/f_V$ can be extracted from the partial
decay width $\Gamma_{V\to e^+e^-}$ by: \be \frac{e}{f_V} =
\left[\frac{3 \Gamma_{V\to e^+e^-}}{2 \alpha_e
|\vec{p_e}|}\right]^{\frac{1}{2}}, \label{fv} \ee where
$|\vec{p_e}|$ is the electron three-vector momentum in the vector
meson rest frame, and $\alpha_e$ is the fine-structure constant.

The following effective Lagrangians are required for vector meson
couplings to the $D$ meson pair and $D^*\bar{D}$:
\bea \nonumber
\mathcal{L}_{V D\bar D} &=& g_{ V  D\bar D}\{D
\partial_\mu {\bar D}-\partial_\mu D\bar{D} \}{V ^\mu},\\\mathcal L_{V D^*\bar D}
&=& -i g_{VD^*\bar D} \varepsilon_{\alpha\beta\mu\nu}
\partial^\alpha V^\beta
\partial^\mu {D^*}^\nu \bar D + h.c.\label{lag-3}
\eea
The effective coupling $ g_{\gamma^* D^* \bar D}(s)$ can then be
expressed in a general form:
\bea g_{\gamma^* D^* \bar D}(s) =
\sum_V \frac{M_V^2}{f_V}\frac{1}{s-M_V^2 + i \sqrt{s}~
\Gamma_V}g_{VD^*\bar D},\label{coupling}
\eea
where $\Gamma_V$ is the total decay width of the vector meson. The
total cross section for $e^+e^- \to D^{*+} D^- +c.c.$ thus reads
\bea
\sigma (e^+e^-\to D^{*+} D^- +c.c.) =\frac{8
\pi}{3}\frac{|\vec{p}|^3}{s^{3/2}}\alpha_e^2\left|\sum_V
\frac{M_V^2}{f_V}\frac{g_{VD^*\bar D}}{s-M_V^2 + i \sqrt{s}~
\Gamma_V}\right|^2. \label{crosssection1}
\eea
where $g_{VD^*\bar D}\equiv g_{VD^{*+}D^-}=g_{VD^{*-}D^+}$. In this
paper, our definition for $g_{VD^*\bar D}$ is different from that in
Ref.~\cite{Zhang:2009gy}. Namely, it does not include the charge
conjugate coupling. In Ref.~\cite{Zhang:2009gy}, $g_{VD^*\bar
D}\equiv \sqrt{2} g_{VD^{*+}D^-}=\sqrt{2} g_{VD^{*-}D^+}$. Thus,
Eq.~(\ref{crosssection1}) has a factor of 2 different from Eq. (12)
in Ref.~\cite{Zhang:2009gy}.

In the following calculation, we mainly consider the contributions
from $\rho, \ \omega$, $J/\psi$, and their radial excitation states.
the contributions from the $\phi$-mesons are dropped because the
$g_{\phi D^* \bar D}$ couplings are strongly suppressed by the
Okubo-Zweig-Iizuka (OZI) rule. The $g_{\Upsilon D^* \bar D}$
couplings are also suppressed by the OZI rule. Moreover, the
$\Upsilon$ states are far away from the $D^* \bar D$ threshold.
Thus, their contributions can be safely neglected. Parameters for
the vector mesons are listed in Table~\ref{tab-1}.

\begin{table}
\caption{ Resonance parameters of the vector mesons adopted in this
study. They are taken from PDG~\cite{pdg2008}.} \label{tab-1}
\begin{tabular}{c|c|c|c|c|c|c|c|c|c|c}\hline
 & $\rho(770)$ & $\rho(1450)$& $\rho(1700)$ &$\omega(782)$&
 $\omega(1420)$& $\omega(1680)$ & $~~~J/\psi~~~$ & $\psi(3686)$ &
 $\psi(3770)$ &$\psi(4040)$ \\\hline
 $M_V$ (GeV) & 0.774 & 1.465 & 1.720 & 0.783 &1.45 & 1.62 & 3.097
 &3.686 & 3.773 & 4.039 \\\hline
 $\Gamma_V$(MeV) & 149.4 & 400.0 & 250.0 & 8.5 & 200.0 & 250.0 & 9.32
 $\times 10^{-2}$ & 0.317 & 27.3 & 80 \\\hline
 $\Gamma_{ee}$ (keV) & 7.04 &-&-& 0.6 &0.46& 0.8&
 5.55&2.38&0.265&0.86\\\hline
\end{tabular}
\end{table}

The asymptotic behavior of $g_{\gamma^* D^* \bar D}(s)$ has been
discussed in Ref.~\cite{Aliev:1994qf}. As $s\to \infty$, form factor
$g_{\gamma^* D^* \bar D}(s)$ must decrease at least as $s^{-2}$ to
avoid the violation of unitary. As a consequence, the following
relations are obtained for the isospin-1 and the isospin-0
components: \be \sum_{\rho_i} \frac{M_{\rho_i}^2}{f_{\rho_i}}
g_{\rho_i D^*\bar D} = 0 \ , \label{relation-1} \ee and \be
\sum_{V(I=0)} \frac{M_{V(I=0)}^2}{f_{V(I=0)}} g_{V(I=0) D^*\bar D} =
0 \ . \ee As discussed in Ref.~\cite{Aliev:1994qf}, the above
asymptotic relation implies that at least two $\rho$ mesons are
needed in the VMD model.

We also adopt the following relations given by the SU(3) quark
model:
\bea
f_{\omega_i} \simeq 3 f_{\rho_i}, ~~~~ m_{\rho_i}^2 \simeq
m_{\omega_i}^2 \ , \label{relation-2}
\eea
where the factor 3 can be tested well by the partial decay widths
for $\omega_i$ and $\rho_i$ via Eq. (\ref{fv}). With the flavor
symmetry, we also have:
\bea g_{\omega_i {D^*}^+ D^-} = - g_{\rho_i {D^*}^+ D^-}.
\label{relation-3}
\eea

It is worth noting that the above relations, i.e. Eqs.
(\ref{relation-1}), (\ref{relation-2}) and (\ref{relation-3}), imply
a negligible contribution from the $\rho$ and  $\omega$ mesons in
the production of $D^* \bar{D}+c.c.$ pairs in $e^+e^-$ annihilation,
although they are dominant in the $D^*$ radiative decays.

The strong coupling $g_{V D^*\bar D}$ for those charmonium states
below the $D^*\bar{D}$ threshold cannot be directly extracted from
the experimental data, such as $g_{J/\psi D^*\bar D}$. Their
coupling values generally have large discrepancies in different
models. The relation between $g_{V D^*\bar D}$ and $g_{V D\bar D}$
can be parameterized by: \be\label{g_v-1} g_{V D^*\bar D} = g_{V
D\bar D}\times g_V, \ee where $g_V$ is a parameter with inverse of
mass dimension. In the heavy quark mass limit
\cite{Deandrea:2003pv}, one has $g_V = 1/\sqrt{m_{D^*} m_D} \simeq
0.52 \ \mbox{GeV}^{-1}$. In contrast, the relativistic potential
model of Ref.~\cite{Colangelo:1994jc} gives:
\be\label{g_v}
g_V =
\frac{e_Q}{\Lambda_Q} + \frac{e_q}{\Lambda_q},
\ee
where $e_Q$ is the heavy quark charge and $e_q$ is the light quark
charges, and the expressions of $\Lambda_Q$ and $\Lambda_q$ can be
found in Ref.~\cite{Colangelo:1994jc}. It should be mentioned that
as pointed out in Ref.~\cite{Casalbuoni:1996pg},
Equation~(\ref{g_v}) is a general consequence of decomposing the
electromagnetic current into a heavy and light part in the radiative
decay $D^*\to D\gamma$. In Ref.~\cite{Oh:2000qr}, the value
$g_{J/\psi} \sim 1.0~ \ {\mbox{GeV}}^{-1}$ is extracted, and QCD sum
rules give $g_{J/\psi} = 0.69 \pm 0.14 \
{\mbox{GeV}}^{-1}$~\cite{Matheus:2003pk}.

In the numerical calculation, we neglect the coupling differences
between the charge and neutral channels for $g_{V D^*\bar D}$ and
$g_{V D\bar D}$. The coupling constants $g_{\psi^\prime D \bar D}$
and $g_{\psi(3770) D\bar D}$ have been discussed in our previous
work~\cite{Zhang:2009gy}, and $g_{\psi(3770) D\bar D} \simeq 12.7$
is extracted from the experimental result~\cite{pdg2008} by the
effective Lagrangian approach. The coupling $g_{\psi^\prime D \bar
D}\simeq 9.05$ is determined by fitting the lineshape of $e^+e^- \to
D\bar D$ process. We adopt $g_{J/\psi D \bar D} = 7.44$ from
Ref.~\cite{Oh:2007ej}, which is obtained by the VMD model.

We must note that the cross section measurement gives access to the
absolute value of $|g_{\gamma^* D^* \bar D}(s)|$. But $g_{\gamma^*
D^* \bar D}(s)$ is a complex function of $s$ in the time-like
region. Moreover, the prescription at the hadronic level will
introduce a phase factor  $e^{i\phi}$ to each resonance amplitude.
These phase angles, apart from an overall phase, can be determined
by fitting the Belle data for $e^+ e^-\to
D^{*+}D^-+c.c.$~\cite{Abe:2006fj}.

As one can see that the Belle data cover a rather high $\sqrt{s}$
region, it is natural to anticipate that the low $\sqrt{s}$ form
factor would be less sensitive to the data constraints. Taking into
account this, we include the real photon data for $D^* \to D \gamma$
in the numerical fitting. In the real photon limit, it is rather
direct to obtain the partial width of $D^* \to D \gamma$ by
\bea
\Gamma(D^*\to D \gamma) = \frac{\alpha_e}{3} g^2_{\gamma D^*D}(0)
q^3, \label{radiate}
\eea
where $q$ is the photon energy in the $D^*$ rest frame. Again,
$g_{\gamma D^*D}(0)$ can be expressed as Eq.~(\ref{coupling}). This
would provide constraints on the parameters from light vector meson
components.

The decay of  $D^* \to D e^+e^-$ also gives access to the transition
form factor at low $\sqrt{s}$. The matrix element of the $D^* \to D
e^+e^-$ decay via single photon transition can be written as
 \bea
T = g_{\gamma^* D^* \bar D}(k^2)  \frac{e^2}{k^2}
\varepsilon_{\alpha\beta\mu\nu} p_D^\beta p_{D^*}^\mu\epsilon^\nu
\bar {u}(k_1) \gamma_\alpha v(k_2) ,\label{ds_to_dee}
\eea
where $k^2$ is the invariant mass of the lepton pair. The formula of
the differential probability  can be described by the following
expression:
\bea\label{two-lepton}
\frac{d \Gamma}{d k^2 d Q^2} &=& \frac{ \alpha_e ^2 |g_{\gamma^* D^*
\bar D}(k^2)|^2}{48 \pi m_{D^*}^3}\left[ k^2 + 2
Q^2 - 2 m_D^2 -2 m_{D^*}^2 \right. \nonumber \\
&& +\frac{(m_{D^*}^2 - m_e^2 -Q^2)^2 + (m_{D}^2 - m_e^2 -Q^2)^2 - 8
m_e^2 Q^2}{k^2}\nonumber  \\ && \left. + \frac{2 (m_{D^*}^2 -
m_D^2)^2 m_e^2}{k^4}\right],
\eea
where $Q^2$ is defined as $Q^2 = (k_2 + p_D)^2$ (or $Q^2 = (k_1 +
p_D)^2$). We should note that this process is strongly suppressed by
an additional EM coupling in respect of $D^*\to D\gamma$. Therefore,
it is relatively difficult to measure this branching ratio in
experiment.

\section{Numerical Results}\label{sec-iii}

Now, we switch to the details of the numerical fitting. First, we
give a brief discussion about the fitting scheme:

(i) Since only the relative phase can be measured in experiment, we
set $\phi_{J/\psi}=0^\circ$, and then the other phase angles are
defined in respect of $\phi_{J/\psi}$. Meanwhile, since the cross
sections are not sensitive to the light vector meson contributions,
the relative phases $\phi_{\rho_i}$ and $\phi_{\omega_i}$ are set
the same and denoted by $\phi_{LV}$ in the fitting.

(ii) As shown by the cross sections around 4.2 GeV, there is no
clear evidence for an enhancement due to the presence of a
resonance. Therefore, the data for $e^+e^- \to D^*\bar D+c.c.$
cannot constrain $\phi_{Y(4260)}$ at all. For simplicity, we set
$\phi_{Y(4260)}=0^\circ$.

(iii) For $\psi(4040)$ and $Y(4260)$, $g_{V D^*\bar D}$ are not
clear. Especially, the $\frac{M_V ^2}{f_V}$ for $Y(4260)$ is also
unavailable. In the numerical fitting, these two couplings are
always combined. Thus, we define $g_V^{\mbox{eff}}\equiv\frac{M_V
^2}{f_V} \times g_{V D^*\bar D}$ for $\psi(4040)$ and $Y(4260)$. In
total, the fitting parameters include the relative phases
$\phi_{LV}$, $\phi_{\psi^\prime}$, $\phi_{\psi(3770)}$,
$\phi_{\psi(4040)}$ , and the couplings
$g_{\psi(4040)}^{\mbox{eff}}$ and $g_{\psi(4260)}^{\mbox{eff}}$.

As discussed previously, the charmed meson couplings to the light
mesons are obtained in the chiral and heavy quark
limits~\cite{Cheng:2004ru}: $g_{\rho_i D^* D } = \sqrt{2} \lambda
m_{\rho_i}/f_\pi$, with $f_\pi = 132$ MeV, and $\lambda = 0.56 \
\mbox{GeV}^{-1}$~\cite{Yan92}. These couplings contain uncertainties
arising from $g_V$ as illustrated by Eq.~(\ref{g_v}). Also, since
the constraints on the light vector mesons are rather weak in the
data for $e^+e^- \to D^* \bar{D} +c.c.$, we thus include the data
for $D^*\to D\gamma$ to constrain couplings $g_\rho$ and $g_\omega$.

For charmonium coupling to the charmed mesons, the following
relation is assumed: \be g_\psi \equiv g_{J/\psi} \simeq
g_{\psi^\prime} \simeq g_{\psi(3770)} \simeq g_{\psi(4040)} \ , \ee
which can be determined by \be \frac{\Gamma(\psi(4040)\to D\bar
D)}{\Gamma(\psi(4040)\to {D^*}\bar D+ c.c.)} =
\frac{|\vec{p_1}|^3}{g_\psi ^2M_{\psi(4040)}^2|\vec{p_2}|^3},
\label{ratio} \ee where $\vec{p_1}$ and $\vec{p_2}$ are the three
momenta of the final charmed mesons in $\psi(4040)\to D\bar D $ and
$ \psi(4040)\to {D^*}\bar D+ c.c.$, respectively. Hence, given the
experimental data~\cite{pdg2008}, \be \frac{\Gamma(\psi(4040)\to
D^0\bar D^0)}{\Gamma(\psi(4040)\to {D^*}^0\bar D^0+ c.c.)} = 0.05
\pm 0.03, \ee we have $g_\psi = 1.73  \pm 0.52 ~\mbox{GeV}^{-1}$ by
taking the average value corresponding to the datum bound. This
value appears to be larger than $g_V \simeq 0.52 \ \mbox{GeV}^{-1}$
extracted by Ref.~\cite{Deandrea:2003pv} and $g_V \sim 1.0~ \
{\mbox{GeV}}^{-1}$ by Ref.~\cite{Oh:2000qr}. In order to examine the
impact of the uncertainties due to $g_V$, we shall fix $g_\psi =
1.73 ~\mbox{GeV}^{-1}$ and $0.52 \ \mbox{GeV}^{-1}$, respectively,
in the numerical fitting.

\subsection{With relatively large $g_\psi$ determined by experimental
data}\label{sub-1}

In Table \ref{tab-2}, all the fitted parameters are listed. It shows
that the coupling $g_{Y(4260)}^{\mbox{eff}}$ has a large
uncertainty, which reflects the negligible role played by $Y(4260)$
in the fitting. Further experimental data with high accuracy are
needed to extract its resonance parameters.

\begin{table}[ht]
\caption{ Model parameters obtained from the $\chi^2$ minimization
fitting with $g_\psi=1.73$ GeV$^{-1}$. Coupling $g_V^{\mbox{eff}}$
is defined by $g_V^{\mbox{eff}}\equiv\frac{M_V ^2}{f_V} \times g_{V
D^*\bar D}$. The phase angles are in radian. The reduced $\chi^2$ is
$\chi^2/\mbox{d.o.f} = 41.2/51$. } \label{tab-2}
\begin{tabular}{|c|c|c|c|c|c|c|} \hline
    parameter   &$\phi_{LV}$       &$\phi_{\psi^\prime}$   &$\phi_{\psi(3770)}$    &$\phi_{\psi(4040)}$    &$g_{\psi(4040)}^{\mbox{eff}}$  &$g_{Y(4260)}^{\mbox{eff}}$ \\ [1ex] \hline
                &$-0.41\pm 0.17$    & $2.36 \pm 0.32$       &$4.54 \pm 0.22$        &$-0.68 \pm 0.21$       &$0.37 \pm 0.05$                &$0.02 \pm 0.03$           \\ [1ex] \hline
\end{tabular}
\end{table}

With the help of Eq.~(\ref{fv}) and the fitted value for
$g_{\psi(4040)}^{\mbox{eff}}$, we obtain  $g_{\psi(4040) D^* \bar D}
= 0.74 \pm 0.1 \ \mbox{GeV} ^{-1}$. Consequently, the partial decay
width of $\psi(4040)\to D^* \bar D + c.c.$ can be accessed: \bea
\nonumber \Gamma(\psi(4040)\to {D^*}^+ D^- + c.c.) &=& 5.1 \pm 1.0 \
\mbox{MeV},
\\ \Gamma(\psi(4040)\to {D^*}^0 \bar{D^0} + c.c.) &=& 5.5 \pm 1.1 \
\mbox{MeV}.
\eea

An interesting result from this fitting is that, although $g_{VD^*
\bar D}$ bares large uncertainties, the excitations of the
charmonium states $J/\psi$, $\psi^\prime$, and $\psi(3770)$, and
their interferences play a major role on the interpretation of the
effective coupling $g_{\gamma^* D^* \bar D}$ or $g_{\gamma^* D \bar
D}$~\cite{Zhang:2009gy}. This feature can be seen more clearly via
the fitted cross sections.

\begin{figure}[ht]
\begin{center}
\begin{tabular}{ccccc}
\scalebox{0.5}{\includegraphics{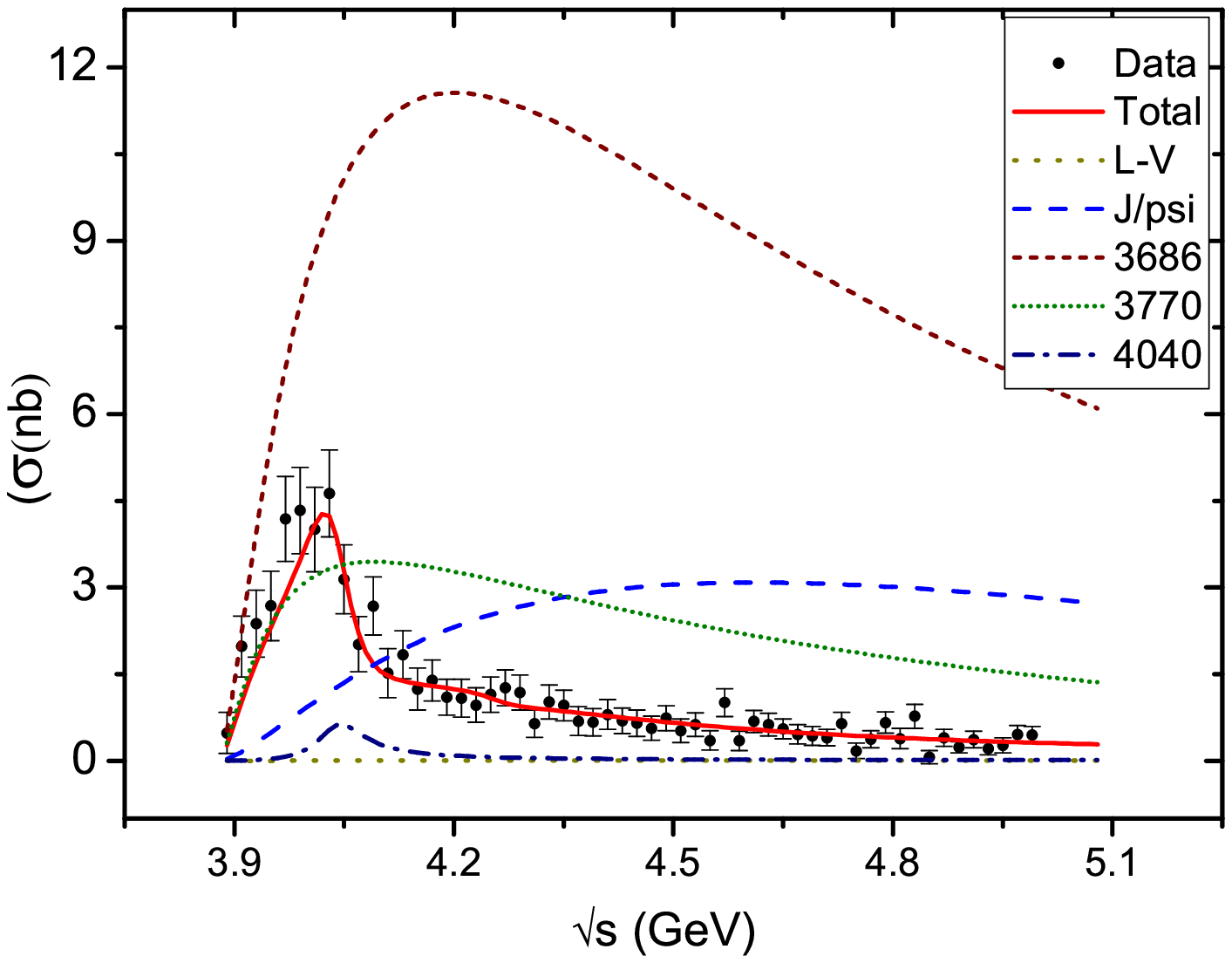}}&&\scalebox{0.5}{\includegraphics{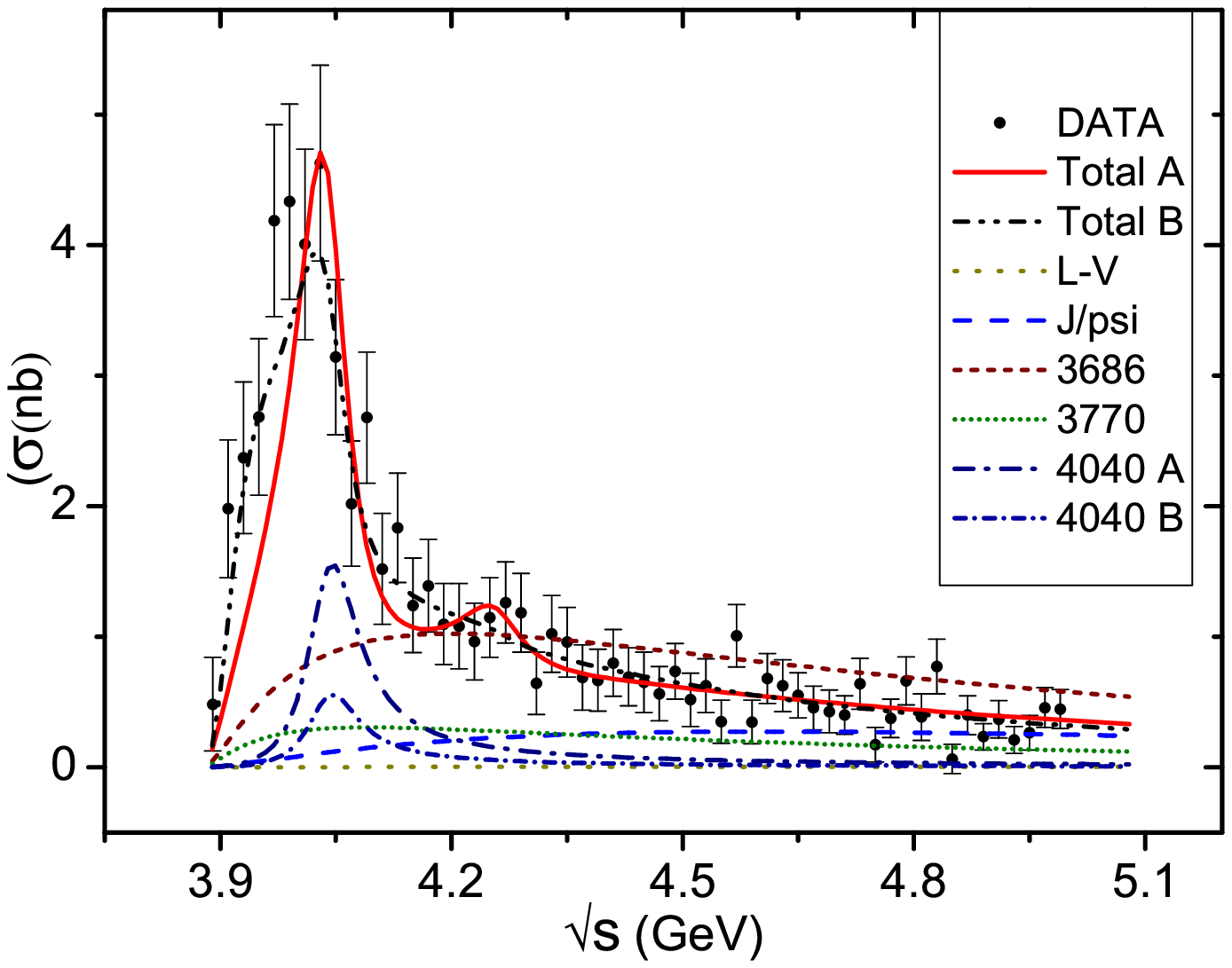}}\\
(a)&&(b)\\
\end{tabular}
\end{center}
\caption{The Belle data for the $e^+e^-\to D^{*+} D^-+c.c.$ cross
section~\protect\cite{Abe:2006fj} are fitted by the $\chi^2$
minimization method with different values for $g_\psi$. Panel (a) is
obtained with $g_\psi=1.73$ GeV$^{-1}$, and panel (b) with
$g_\psi=0.52$ GeV$^{-1}$. The dotted, dashed, short-dashed and
short-dotted lines are for exclusive contributions from the
light-vector mesons (LV), $J/\psi$, $\psi^\prime$ and $\psi(3770)$,
respectively. In panel (a), the solid line represents the overall
results, while the dash-dotted line if for exclusive contribution
from $\psi(4040)$. In panel (b), the solid, dash-dot-dotted lines
are for the overall results from two fitting schemes, i.e. Scheme-A
and Scheme-B, respectively, while the dash-dotted and
short-dash-dotted are for contributions from $\psi(4040)$ in these
two schemes. } \label{fig-2}
\end{figure}

In Fig.~\ref{fig-2}(a), the total cross section and cross sections
for exclusive resonances are plotted. We do not show the curve of
$Y(4260)$ since its contribution is negligibly small. Interestingly,
other charmonia, such as $J/\psi$, $\psi^\prime$, and $\psi(3770)$,
have large exclusive cross sections. In particular, the cross
section for the $\psi^\prime$ excitation over-shoots the data
apparently, and cancellations among these three amplitudes are
required to reproduce the lineshape of the cross sections.

\begin{figure}[ht]
\begin{center}
\begin{tabular}{ccccc}
\scalebox{0.5}{\includegraphics{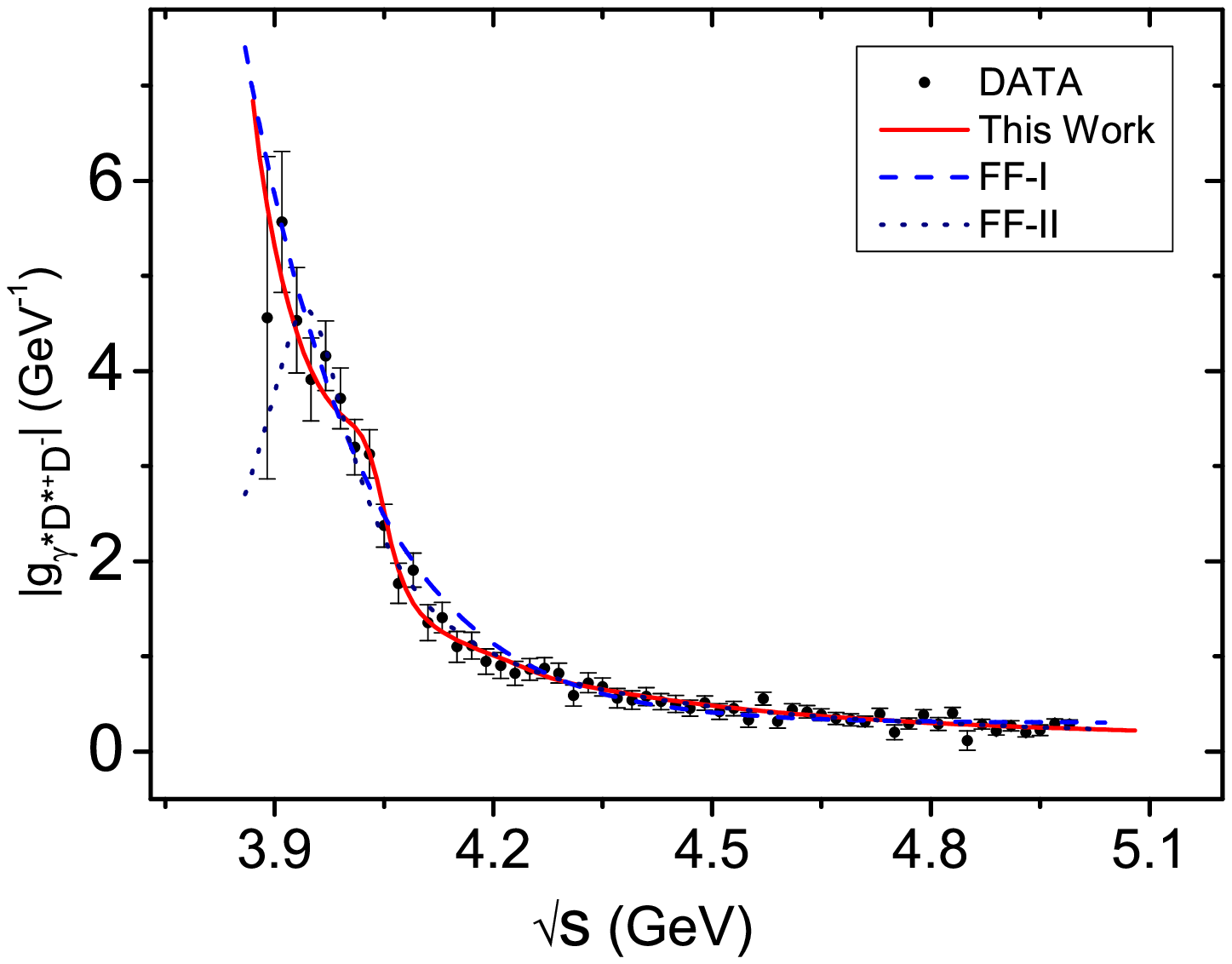}}&&\scalebox{0.5}{\includegraphics{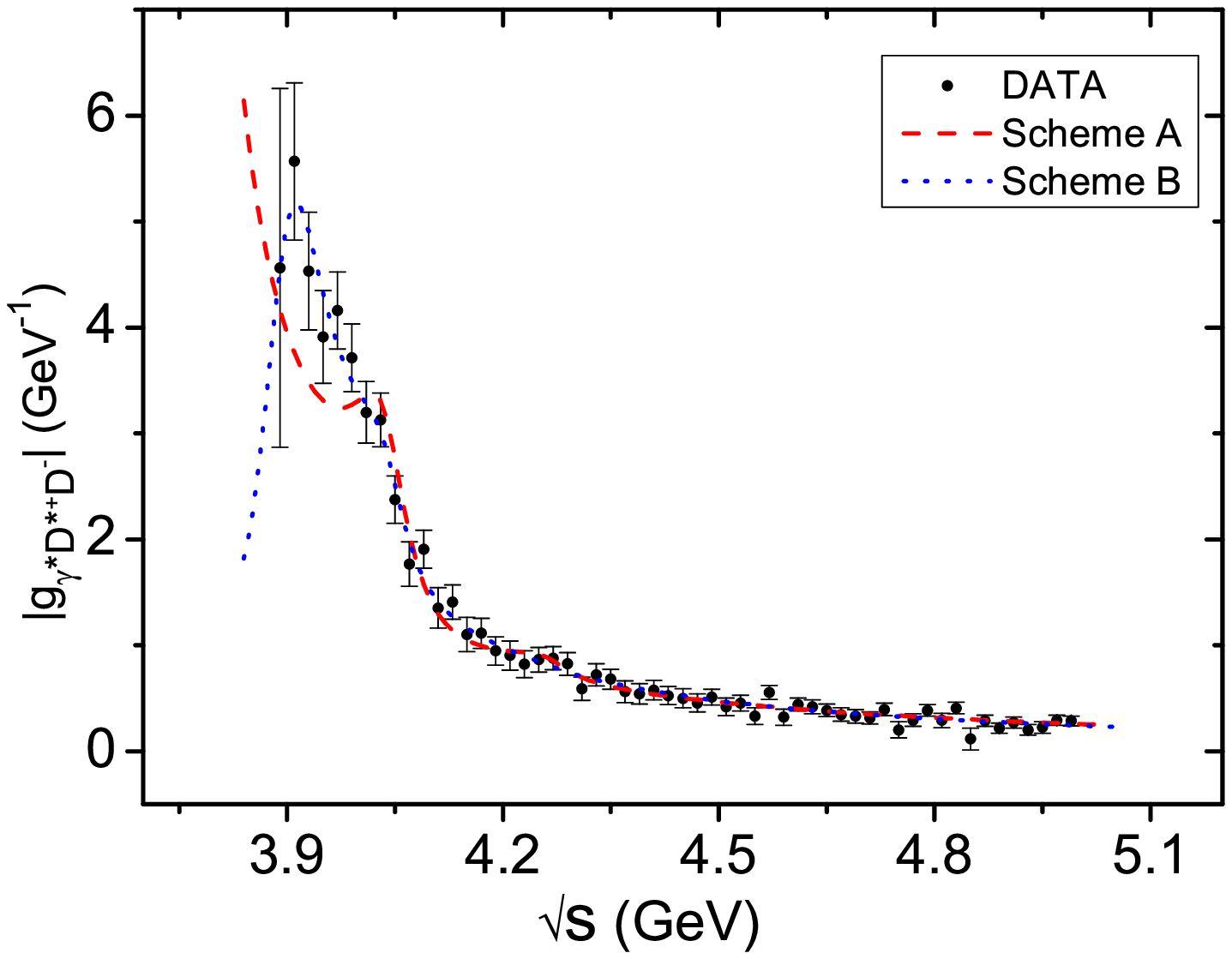}}\\
(a)&&(b)\\
\end{tabular}
\end{center}
\caption{The $\sqrt{s}$-dependence of form factor $|g_{\gamma^*
D^{*+} D^-}(s)|$ extracted from the fitting results. In panel (a),
the solid line is obtained from the fitting results with
$g_\psi=1.73$ GeV$^{-1}$, while the dashed and dotted lines are for
FF-I and FF-II results from Ref.~\cite{Zhang:2009gy}. In panel(b),
the dashed and the dotted lines are obtained from the fitting
results with $g_\psi=0.52$ GeV$^{-1}$ for Scheme-A and B,
respectively.} \label{fig-3}
\end{figure}

We can then extract the  form-factor $|g_{\gamma^* D^{*+} D^-}|$ in
the whole time-like region. In Fig.~\ref{fig-3}(a), we first look at
the region around the $D^*\bar D$ threshold up to 5.0 GeV. As
demonstrated by the solid line, the data can be described perfectly.
We also include those two empirical fits presented in
Ref.~\cite{Zhang:2009gy} as a comparison. The dashed line is
generated by fitting the form factor data with an exponential
function, i.e. with form factor one (FF-I):
\begin{equation}
\sqrt{2} g_{\gamma^*  D^{*+} D^- }(s) = g_1
\exp{[-(s-(m_D+m_{D^*})^2)/t_1]} + g_0 \ ,
\end{equation}
where $x=0$ corresponds to the $D^*\bar{D}+c.c.$ threshold, and
$g_1$, $t_1$, and $g_0$ are fitting parameters. The dotted line is
given by fitting the data with a single resonance:
\begin{eqnarray}
\sqrt{2} g_{\gamma^*  D^{*+} D^-}(s) = \left|\frac{b_0}{s - m_X ^2 +
i m_X \Gamma _X} + b_1\right| \ ,
\end{eqnarray}
with a background term $b_1$. The parameter $b_0$ can be regarded as
the product of the $\gamma^* X$ coupling and $XD\bar{D^*}$ coupling.
This parametrization agrees with the data at higher energies, but
drops at the threshold. In the above two equations, a factor
$\sqrt{2}$ has been included for the change of conventions here. All
the parameters have been given in Ref.~\cite{Zhang:2009gy}.

\begin{figure}[ht]
\begin{center}
\begin{tabular}{ccccc}
\scalebox{0.5}{\includegraphics{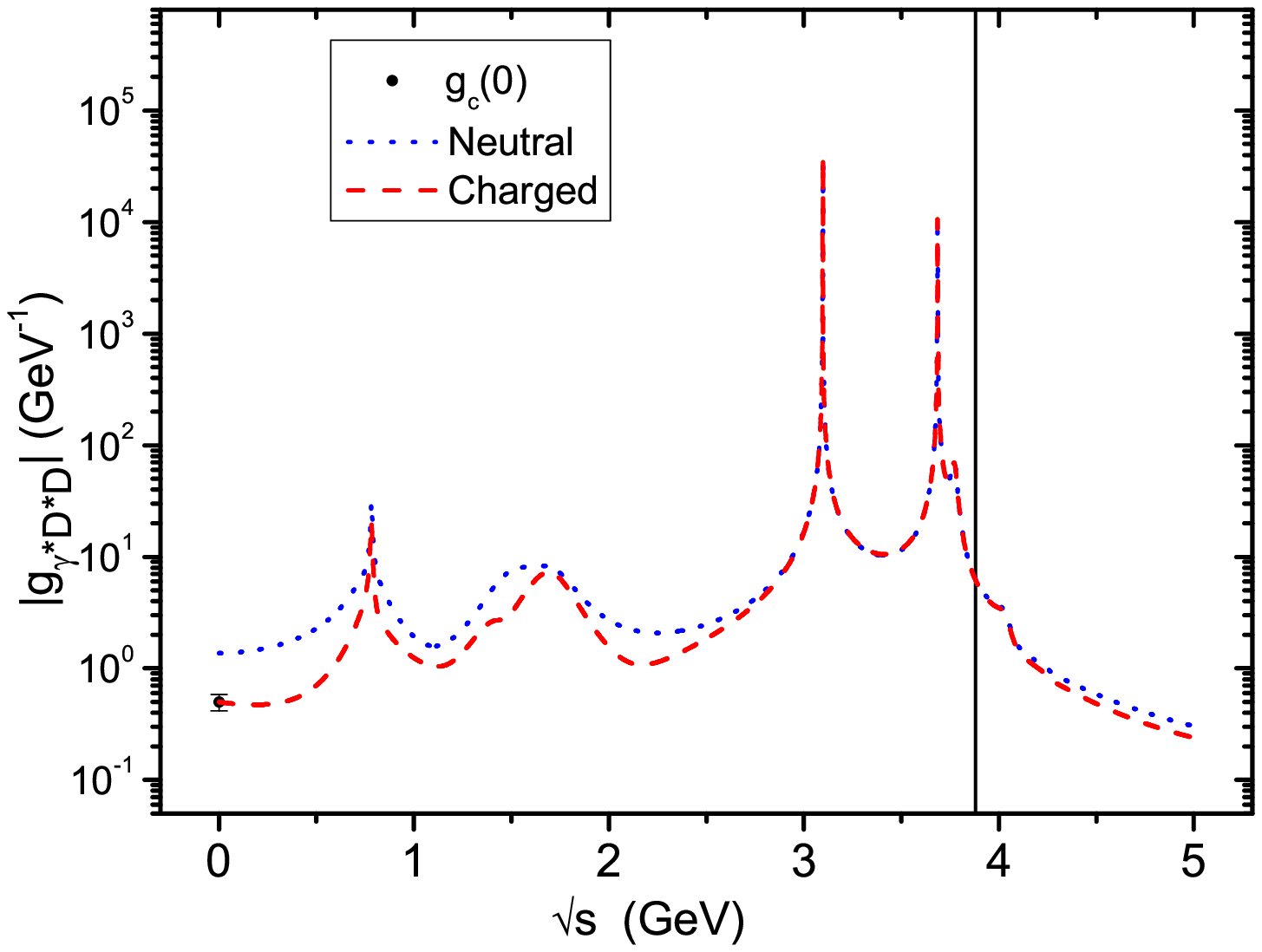}}&&\scalebox{0.5}{\includegraphics{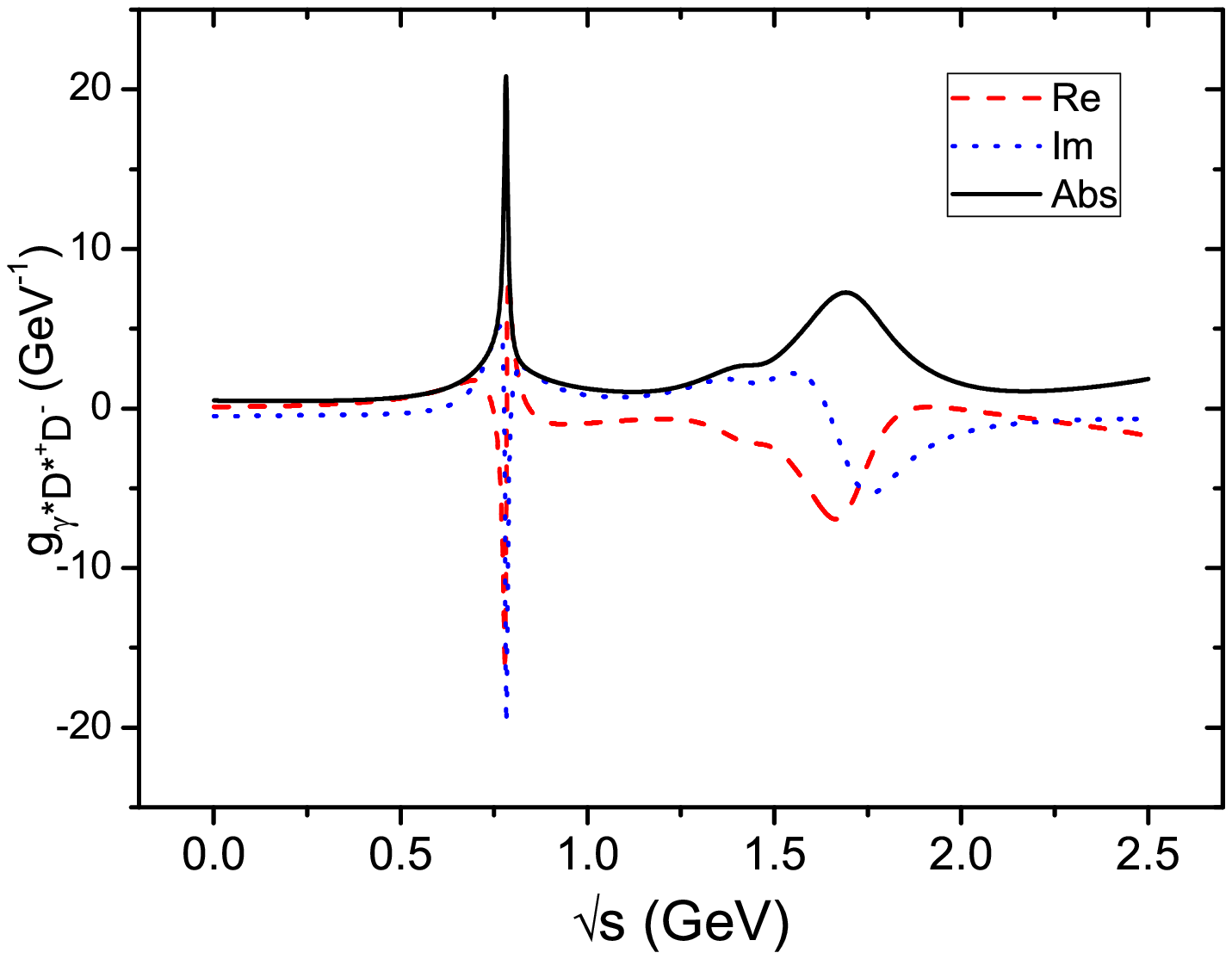}}\\
(a)&&(b)\\
\end{tabular}
\end{center}
\caption{The $\sqrt{s}$-dependence of form factor $|g_{\gamma^*
D^*\bar D}(s)|$ extracted from the fitting results with
$g_\psi=1.73$ GeV$^{-1}$. In panel (a), the dashed and dotted line
stand for the form factors for the charged and neutral channel,
respectively. In panel (b), the solid line is the form factor
$|g_{\gamma^* D^{*+} D^-}(s)|$ for the charged channel, while the
dashed and dotted line denote respectively the real and imaginary
part of $g_{\gamma^* D^{*+} D^-}(s)$. }\label{fig-4}
\end{figure}

In Fig.~\ref{fig-4}(a), we plot the $\sqrt{s}$-dependence of the
form factor $g_{\gamma^* D^*\bar D}(s)$ for both charged and neutral
channel in the whole time-like region up to 5.0 GeV. Since the
coupling $g_{\psi^\prime D^*\bar{D}}$ (and $g_{J/\psi D^*\bar{D}}$)
has the same value in these two channels, these two form factors
converge to each other around the threshold region. In the
low-$\sqrt{s}$ region, the discrepancy arises from the total width
difference between $D^{*\pm}\to D^\pm \gamma$ and $D^{*0}\to
D^0\gamma$. For the latter, the experimental data only give an upper
limit, i.e. $\Gamma(D^{*0}\to D^0\gamma)<2.1$ MeV~\cite{pdg2008}.
This corresponds to $|g_{\gamma D^{0*} D^0}(0)| < 11.3
~\mbox{GeV}^{-1}$, which has not been marked in the figure.

In Fig.~\ref{fig-4}(b), the form factor in the small $\sqrt{s}$
region is plotted in association with its real and imaginary part.
The resonance structures from $\rho(770)$, $\rho(1450)$ and
$\rho(1700)$ are distinguishable. However, it should be cautioned
that this kinematic region would suffer from a lack of information
about the light vector meson couplings to $D^*\bar{D}$. Our model
bridges the real photon form factor with the high-$\sqrt{s}$ one,
but inevitably leaves the middle range with large uncertainties. We
drop the high-$\sqrt{s}$ part between 2.5 GeV and 4.5 GeV since the
structure of the real and imaginary part appears trivially as either
very narrow peaks or very narrow dips.

The following points are advocated to understanding this fitting
results:

(i) Numerically, the large contributions from $\psi^\prime$ are due
to a relatively large $g_\psi$ as shown by Eq.~(\ref{g_v-1}). In
Ref.~\cite{Zhang:2009gy}, we showed that the coupling
$g_{\psi^\prime D\bar{D}}$ can be well-constrained by the cross
section lineshape of $e^+ e^-\to D\bar{D}$. Thus, the coupling
$g_{\psi^\prime D^*\bar{D}}$ is actually enhanced by $g_\psi$ via
Eq.~(\ref{g_v-1}).

(ii) The relative phases appear to be sensitive to the cross section
lineshape of $e^+e^- \to D^* \bar D$, and cancellations among the
dominant amplitudes seem to be inevitable. Such interferences
generally would affect the extraction of resonance parameters.
Because of this, it is desirable to have a precise measurement of
the lineshape of $e^+e^- \to D^* \bar{D}+c.c.$.


In order to clarify the interferences among the charmonium states,
we identify the exclusive contributions from light vector mesons and
charmonium states in $D^{*\pm}\to D^\pm \gamma$, where we expect
that the light vector mesons should play a significant role. We
first inspect the separated contributions to the partial width of
$D^{*\pm}\to D^\pm \gamma$ from the light vector mesons and
charmonia, and the results are as follows:
\bea \nonumber
\Gamma_{LV}(D^{*\pm} \to
D^{\pm}\gamma ) &=& 2.21 \  \mbox{keV}, \\
\Gamma_{\psi}(D^{*\pm} \to D^{\pm}\gamma ) &=& 1.61 \ \mbox{keV},
\eea
where $\Gamma_{LV}$ and $\Gamma_{\psi}$ are decay widths contributed
by the light vector mesons, i.e. $\rho, \omega$ etc, and the
charmonia, i.e. $J/\psi$, $\psi^\prime$ etc. Interestingly, it shows
that, although the light vector mesons play a dominant role,
contributions from the charmonium states are still sizeable. In
comparison with the experimental result $\Gamma({D^*}^{\pm} \to
D^{\pm}\gamma) = 1.54 ~\mbox{keV}$~\cite{pdg2008}, one can see that
a destructive interference between these two components is required.

Such destructive phases are also present within the charmonium
states. As follows, we list the exclusive contributions from
$J/\psi$, $\psi^\prime$, and $\psi(3770)$:
\bea
\Gamma_{J/\psi}({D^*}^{\pm} \to  D^{\pm}\gamma) &=& 8.07 \
\mbox{keV}, \nonumber\\
\Gamma_{\psi^\prime}({D^*}^{\pm} \to
 D^{\pm}\gamma) &=& 5.11 \ \mbox{keV}, \nonumber\\
\Gamma_{\psi(3770)}({D^*}^{\pm} \to  D^{\pm}\gamma) &=& 0.93 \
\mbox{keV}.
\eea
Other charmonium exclusive contributions are negligibly small. So we
do not list them here.

\subsection{With smaller $g_\psi$ by heavy quark theory}
As pointed out earlier, the relatively large contributions from the
charmonium states, in particular, $J/\psi$ and $\psi^\prime$, are
mainly due to the larger value of $g_\psi$ in comparison with those
given by the heavy quark theory~\cite{Deandrea:2003pv} and the
relativistic potential model~\cite{Colangelo:1994jc}. In order to
examine the impact from a possible overestimate of the coupling
$g_\psi$ due to the experimental uncertainties~\cite{pdg2008}, we
perform another fit of the $e^+e^-\to D^{*+} D^-+c.c.$ cross
sections adopting $g_\psi=0.52$ GeV$^{-1}$~\cite{Deandrea:2003pv}.

Two fitting schemes are considered. In Scheme-A, we fix
$g_\psi=0.52$ GeV$^{-1}$ and fit the data with the same parameters.
The fitted parameters are listed in Table~\ref{tab-4}. Relatively
large $\chi^2$ is found and the fit cannot account for the cross
section lineshape near threshold as shown by the solid line in
Fig.~\ref{fig-3}(b). The major deviations occur near threshold where
a cross section deficit is revealed. The relatively small $g_\psi$
also leads to suppressed contributions from $J/\psi$, $\psi^\prime$
and $\psi(3770)$. In contrast with the fitting of
Subsection~\ref{sub-1}, this scheme requires constructive phases
among the charmonium amplitudes.

In Scheme-B, we still fix $g_\psi=0.52$ GeV$^{-1}$, and introduce
contributions from an additional resonance $X(3900)$ in order to
overcome the cross section deficit near threshold. When we introduce
the X(3900) resonance, it is difficult to obtain the fitting
solution if we leave the resonance parameters free, i.e. the mass
and total width. We then take $m_X= 3.9 \ \mbox{GeV}$ and $\Gamma_X
= 89.8 \ \mbox{MeV}$, which were extracted from $e^+ e^- \to
D\bar{D}$ in Ref.~\cite{Zhang:2009gy}, as an input. The other fitted
parameters are listed in Table~\ref{tab-4}, and relatively smaller
$\chi^2$ is found. In Fig.~\ref{fig-2}(b), the results for Scheme-A
and B are compared with each other. It shows that contributions from
the $\psi(4040)$ are relatively suppressed due to the presence of
the $X(3900)$. Some interfering effects for the $Y(4260)$ are also
observed in Scheme-A, but not in Scheme-B. As shown in
Table~\ref{tab-4}, the combined coupling for the $Y(4260)$ still has
large uncertainties.

\begin{table}[ht]
\caption{ Model parameters obtained from the $\chi^2$ minimization
fitting with $g_\psi=0.52$ GeV$^{-1}$. We fix $m_X= 3.9 \
\mbox{GeV}$ and $\Gamma_X = 89.8 \ \mbox{MeV}$, which are from
Ref.~\cite{Zhang:2009gy}. The phase angles are in radian. }
\label{tab-3}
\begin{tabular}{|c|c|c|c|c|c|c|c|c|c|} \hline
    parameter   &$\phi_{LV}$       &$\phi_{\psi^\prime}$   &$\phi_{\psi(3770)}$    &$\phi_{\psi(4040)}$    &$g_{\psi(4040)}^{\mbox{eff}}$  &$g_{Y(4260)}^{\mbox{eff}}$  &$g_{X(3900)}^{\mbox{eff}}$  &$\phi_{X(3900)}$  & $\chi^2/\mbox{d.o.f}$\\ [1ex] \hline
    Scheme A    &$5.72\pm 0.63$    & $3.56 \pm 0.10$        &$3.69 \pm 0.22$        &$0.0  \pm 0.11$        &$0.58 \pm 0.05$                &$0.08 \pm 0.03$             &  -                         & -       & 63.9/51\\ [1ex] \hline
    Scheme B    &$5.69\pm 0.66$    & $4.21 \pm 0.43$        &$2.23 \pm 0.35$        &$5.62 \pm 0.27$        &$0.35 \pm 0.07$                &$0.01 \pm 0.04$             &  $1.89 \pm 0.44$           & $3.03\pm0.26$   &38.8/49       \\ [1ex] \hline
\end{tabular}
\end{table}

In Fig.~\ref{fig-3}(b), the form factors extracted from Scheme-A and
B are compared with the data. Again, we see that a smaller value for
$g_\psi$ cannot account for the form factor near threshold. Although
the inclusion of an additional resonance $X(3900)$ can optimize the
description, the form factor appears to drop quickly in the
subthreshold region. This will lead to different trends of the form
factor in the middle range of kinematics, i.e. $\sqrt{s}$ is between
$2\sim 3.9$ GeV.

The results with a smaller value for $g_\psi$ seem to be different
from those with a larger value. The following points can be learned
and conjectured:

i) The resonance parameters extracted from the cross sections would
contain uncertainties inevitably due to the uncertainty with
$g_\psi$.

ii) The need of $X(3900)$ with a small $g_\psi$ would bring
questions on the underlying physics. On the one hand, if such a
resonance indeed exists, problem will arise from how to organize it
within the quark model framework. The systematic study of quark
potential model seems not to have a place for this state below 4.2
GeV~\cite{Barnes:2005pb}.

iii) To void the conflicts with the quark model, one possibility for
such a structure would be due to the open channel effects of
$D_s^*\bar{D_s}+c.c.$ Its threshold is 4.08 GeV, which is not far
away from the $D^*\bar{D}+c.c.$ threshold. The final state
interaction $D_s^* \bar{D}\to D\bar{D^*}$ via kaon exchange can
produce a resonance-like enhancement near threshold. This is similar
to the mechanism discussed in Ref.~\cite{Zhang:2009gy}. Also,
justification of such a possibility would need the data for $e^+
e^-\to D_s^*\bar{D_s}+c.c.$, which unfortunately are unavailable.

iv) As a comparison with the results from Subsection~\ref{sub-1}, we
also list the exclusive contributions of the charmonium components
to the radiative width in Table~\ref{tab-4}. Again, we see that with
the smaller $g_\psi$, the charmonium contributions are also strongly
suppressed in the real photon form factor. Nevertheless, a
destructive phase exists between the $J/\psi$ and $\psi^\prime$
amplitudes. Note that the contributions from the light vector mesons
are unchanged.

\begin{table}[ht]
\caption{ Exclusive contributions to the ${D^*}^{\pm} \to
D^{\pm}\gamma$ partial width from individual charmonium states with
$g_\psi=0.52$ GeV$^{-1}$. $\Gamma_\psi$ is the width given by a
coherent sum of all the charmonium amplitudes.} \label{tab-4}
\begin{tabular}{|c|c|c|c|c|c|} \hline
    Partial width (keV)    &$\Gamma_{J/\psi} $       &$\Gamma_{\psi^\prime}$   &$\Gamma_{\psi(3770)} $    &$\Gamma_{\psi(4040)} $  &$\Gamma_\psi $\\ [1ex] \hline
    Scheme-A    &$0.72$    & $0.46$        &$0.08 $        &$0.01  $       & 0.19\\ [1ex] \hline
    Scheme-B    &$0.72$    & $0.46$        &$0.08 $        &$0.00  $       & 0.31 \\ [1ex] \hline
\end{tabular}
\end{table}

\subsection{Predictions for $D^*\to D e^+ e^-$}

The ambiguity with the charmonium couplings will not affect the
calculations for $D^*\to D e^+ e^-$ as long as the radiative decay
$D^*\to D\gamma$ is fixed. This is simply because the mass of the
electron (positron) is very small, and $\sqrt{s}$ of the virtual
photon is close to the real photon limit.

With the form factor determined in Subsection~\ref{sub-1}, we can
predict the partial decay width of $D^* \to D e^+e^-$ with the help
of Eq.~(\ref{two-lepton}):
\be
\Gamma({D^*}^{\pm} \to D^{\pm} e^+e^-) =
9.95~ \mbox{eV}.
\ee
It corresponds to a branching ratio, $BR(D^{*\pm}\to D^\pm e^+ e^-)
= 1.04\times 10^{-4}$, and should be accessible in experiment, e.g.
at BES-III~\cite{bes-iii}. This decay channel may provide some
further constraints on the role played by the light vector mesons
and charmonium states.

\section{Summary and discussion}\label{sec-iv}

In this work, we study the $D^* \bar D+c.c.$ production in $e^+e^-$
annihilation from the threshold to $\sqrt{s} \simeq 5.0$ GeV in the
VMD model. The recent experimental data from Belle~\cite{Abe:2006fj}
allow us to extract the form factor $g_{\gamma^* D^*\bar{D}}(s)$ in
the time-like region.

Due to the uncertainties with the charmonium couplings to
$D^*\bar{D}+c.c.$, we find two different solutions in the
interpretation of the experimental data: i) With a relatively large
coupling for $\psi D^*\bar{D}$, significantly large contributions
from individual charmonium states are found. Destructive
interferences among those charmonium components are hence required
to bring down the overall cross sections, and then account for the
cross section lineshape. ii) With a relatively small value for the
$\psi D^*\bar{D}$ coupling, i.e. $g_\psi=0.52$
GeV$^{-1}$~\cite{Deandrea:2003pv}, we find an apparent cross section
deficit near threshold, and contributions from other mechanisms are
needed.

We also try to fit the $e^+e^- \to D^{*+} D^- + c.c.$ with $g_\psi =
0.69 \ \mbox{GeV}^{-1}$, which is obtained by QCD sum
rules~\cite{Matheus:2003pk}. It shows that a cross section deficit
still exists near threshold, although is becomes smaller than that
with $g_\psi=0.52$ GeV$^{-1}$, and $\chi^2/\mbox{d.o.f} = 54.6/51$
is found. This suggests that a better fit of the near-threshold
cross section would require a relatively large value for $g_\psi$.
The consequence, however, is that destructive inferences among the
resonances beyond the threshold region would be expected. In case
that $g_\psi$ has a relatively small value, such a cross section
deficit, on the one hand, might imply the presence of an additional
resonance $X(3900)$. On the other hand, we point out that an
enhancement like that could be produced by the $D_s^*\bar{D_s}+c.c.$
open channel effects, and further experimental data will be able to
clarify this issue.

In Ref.~\cite{Zhang:2009gy}, $g_{\gamma^* D^{*+} D^-}(s)$ is
extracted by two simple functions from which the behavior of the
form factor below the $D^*\bar{D}$ threshold is, in principle,
unknown. In this work, our model provides a description of the form
factor $g_{\gamma^* D^*\bar{D}}(s)$ in the interplay region between
the real photon energy and $D^*\bar{D}$ threshold, and some insights
into the evolution of the vector meson contributions can be gained.
Since this is a kinematic region which cannot be directly accessed
by experiment, we expect further investigation of alternative
processes would provide a test of our model, and more information on
the underlying dynamics could be extracted. In particular, the
partial decay width for $D^*\to D e^+e^-$ is predicted in this
framework. With a branching ratio of $BR(D^{*\pm}\to D^\pm e^+
e^-)=1.04\times 10^{-4}$, this channel can be measured by BES-III in
experiment. Our results also support such an ideal that the
enhancement around 3.9 GeV in $e^+ e^-\to
D\bar{D}$~\cite{Pakhlova:2008zza} is caused by the open $D^*\bar{D}$
threshold effects~\cite{Zhang:2009gy}.

It is worth mentioning a recent study of multiple solutions in
extracting physics information from experiment~\cite{Yuan:2009gd}.
Since it is not possible that all the numerical solutions are
correct, a conjecture of selecting the physical solution is that the
physical solution would correspond to the minimal magnitudes of the
amplitudes. Further theoretical investigation of the $\psi
D^*\bar{D}$ couplings and experimental information for $e^+ e^-\to
D_s^*\bar{D_s}+c.c.$ may help disentangle the underlying mechanism
and also test the idea of Ref.~\cite{Yuan:2009gd}.

\section*{Acknowledgement}

We would like to thank G. Pakhlova, C.Z. Yuan and X.L. Wang for
useful discussions regarding the Belle results. This work is
supported, in part, by the National Natural Science Foundation of
China (Grants No. 10675131), Chinese Academy of Sciences
(KJCX3-SYW-N2), and Ministry of Science and Technology of China
(2009CB825200).


\begin{thebibliography}{99}


\bibitem{Abe:2006fj}
  K.~Abe {\it et al.}  [Belle Collaboration],
  Phys.\ Rev.\ Lett.\  {\bf 98}, 092001 (2007)
  [arXiv:hep-ex/0608018].


\bibitem{:2009xs}
  B.~Aubert {\it et al.}  [BABAR Collaboration],
  Phys.\ Rev.\  D {\bf 79}, 092001 (2009)
  [arXiv:0903.1597 [hep-ex]].

\bibitem{pdg2008}
  C.~Amsler {\it et al.}  [Particle Data Group],
  Phys.\ Lett.\  B {\bf 667}, 1 (2008).

\bibitem{Bauer:1977iq}
  T.~H.~Bauer, R.~D.~Spital, D.~R.~Yennie and F.~M.~Pipkin,
  Rev.\ Mod.\ Phys.\  {\bf 50}, 261 (1978)
  [Erratum-ibid.\  {\bf 51}, 407 (1979)].

\bibitem{Bauer:1975bw}
  T.~Bauer and D.~R.~Yennie,
  Phys.\ Lett.\  B {\bf 60}, 169 (1976).

\bibitem{Aliev:1994qf}
  T.~M.~Aliev, E.~Iltan, N.~K.~Pak and M.~P.~Rekalo,
  Z.\ Phys.\  C {\bf 64}, 683 (1994).

\bibitem{Zhang:2009gy}
  Y.~J.~Zhang and Q.~Zhao,
  Phys.\ Rev.\  D {\bf 81}, 034011 (2010)
  [arXiv:0911.5651 [hep-ph]].


\bibitem{Deandrea:2003pv}
  A.~Deandrea, G.~Nardulli and A.~D.~Polosa,
  Phys.\ Rev.\  D {\bf 68}, 034002 (2003)
  [arXiv:hep-ph/0302273].

\bibitem{Colangelo:1994jc}
  P.~Colangelo, F.~De Fazio and G.~Nardulli,
  Phys.\ Lett.\  B {\bf 334}, 175 (1994)
  [arXiv:hep-ph/9406320].

\bibitem{Casalbuoni:1996pg}
  R.~Casalbuoni, A.~Deandrea, N.~Di Bartolomeo, R.~Gatto, F.~Feruglio and G.~Nardulli,
  Phys.\ Rept.\  {\bf 281}, 145 (1997)
  [arXiv:hep-ph/9605342].

\bibitem{Oh:2000qr}
  Y.~s.~Oh, T.~Song and S.~H.~Lee,
  Phys.\ Rev.\  C {\bf 63}, 034901 (2001)
  [arXiv:nucl-th/0010064].

\bibitem{Matheus:2003pk}
  R.~D.~Matheus, F.~S.~Navarra, M.~Nielsen and R.~Rodrigues da Silva,
  arXiv:hep-ph/0310280.



\bibitem{Oh:2007ej}
  Y.~Oh, W.~Liu and C.~M.~Ko,
  Phys.\ Rev.\  C {\bf 75}, 064903 (2007)
  [arXiv:nucl-th/0702077].

\bibitem{Cheng:2004ru}
  H.~Y.~Cheng, C.~K.~Chua and A.~Soni,
  Phys.\ Rev.\  D {\bf 71}, 014030 (2005)
  [arXiv:hep-ph/0409317].

\bibitem{Yan92} T. M. Yan, H. Y. Cheng, C. Y. Cheung, G. L. Lin, Y. C.
        Lin, and H. L. Yu, Phys. Rev. D {\bf 46}, 1148 (1992);
        {\bf 55}, 5851(E) (1997);
         M. B. Wise, Phys.
        Rev. D {\bf 45}, R2188 (1992); G. Burdman and J. Donoghue,
        Phys. Lett. B {\bf 280}, 287 (1992).

\bibitem{Barnes:2005pb}
  T.~Barnes, S.~Godfrey and E.~S.~Swanson,
  Phys.\ Rev.\  D {\bf 72}, 054026 (2005)
  [arXiv:hep-ph/0505002].

\bibitem{bes-iii} D. M. Asner et al, ``Physics at BES-III¡±, Edited by K.T. Chao
and Y.F. Wang, Int. J. of Mod. Phys. {\bf A 24} Supplement 1, (2009)
[arXiv:0809.1869].

\bibitem{Pakhlova:2008zza}
  G.~Pakhlova {\it et al.}  [Belle Collaboration],
  Phys.\ Rev.\  D {\bf 77}, 011103 (2008)
  [arXiv:0708.0082 [hep-ex]].


\bibitem{Yuan:2009gd}
  C.~Z.~Yuan, X.~H.~Mo and P.~Wang,
  arXiv:0911.4791 [hep-ph].

\end{thebibliography}
\end{document}